\documentclass[slac_one]{revtex4}
\usepackage{graphicx}
\usepackage{fancyhdr,amsmath,psfrag,epsfig}
\newcommand{\msusy}{M_{\mathrm{SUSY}}}
\newcommand{\cp}{{\cal CP}}
\newcommand{\twol}{two-loop}
\newcommand{\KL}{\left(}
\newcommand{\KR}{\right)}
\newcommand{\lsim}
{\;\raisebox{-.3em}{$\stackrel{\displaystyle <}{\sim}$}\;}
\newcommand{\gsim}
{\;\raisebox{-.3em}{$\stackrel{\displaystyle >}{\sim}$}\;}
\def\mathswitchr#1{\relax\ifmmode{\mathrm{#1}}\else$\mathrm{#1}$\fi}
\newcommand{\cw}{\mathswitch {c_{\scrs\PW}}}
\newcommand{\sweff}{\sin^2 \theta_{\mathrm{eff}}}
\newcommand{\fehto}{{\em FeynHiggs2.2}}
\newcommand{\fea}{{\em FeynArts}}
\newcommand{\foc}{{\em FormCalc}}
\def\reffi#1{\mbox{Fig.~\ref{#1}}}
\def\la{\lambda}

\def\Si{\Sigma}
\def\de{\delta}
\def\order#1{${\cal O}\KL #1 \KR$}

\def\tiu{\tilde u}
\def\tid{\tilde d}
\def\tiq{\tilde q}

\def\De{\Delta}
\newcommand{\Sstr}{\tilde{s}}
\newcommand{\Scha}{\tilde{c}}

\newcommand{\Sbot}{\tilde{b}}
\newcommand{\Stop}{\tilde{t}}

\newcommand{\StopL}{\tilde{t}_L}
\newcommand{\StopR}{\tilde{t}_R}

\newcommand{\SchaL}{\tilde{c}_L}
\newcommand{\SchaR}{\tilde{c}_R}

\newcommand{\nn}{\nonumber}

\def\slash#1{\setbox0=\hbox{$#1$}#1\hskip-\wd0\dimen0=5pt\advance
       \dimen0 by-\ht0\advance\dimen0 by\dp0\lower0.5\dimen0\hbox
         to\wd0{\hss\sl/\/\hss}}

\newcommand{\gev}{\unskip\,\mathrm{GeV}}

\newcommand{\mev}{\unskip\,\mathrm{MeV}}
\newcommand{\TeV}{\unskip\,\mathrm{TeV}}
\newcommand{\tev}{\unskip\,\mathrm{TeV}}

\def\mathswitch#1{\relax\ifmmode#1\else$#1$\fi}

\newcommand{\MW}{\mathswitch {M_W}}

\newcommand{\MZ}{\mathswitch {M_Z}}

\newcommand{\mc}{\mathswitch {m_c}}
\newcommand{\mb}{\mathswitch {m_b}}
\newcommand{\mt}{\mathswitch {m_t}}
\newcommand{\mq}{\mathswitch {m_q}}
\def\al{\alpha}
\newcommand{\aeff}{\al_{\mathrm{eff}}}
\newcommand{\scrs}{\scriptscriptstyle}
\newcommand{\sw}{\mathswitch {s_{\scrs\PW}}}
\newcommand{\PW}{\mathswitchr W}
\newcommand{\CZb}{\cos 2\beta\hspace{1mm}}

\newcommand{\mhmax}{\mh^{\rm max}}

\newcommand{\mh}{\mathswitch {m_h}}

\newcommand{\MA}{\mathswitch {M_A}}
\newcommand{\onel}{one-loop}
\newcommand{\BC}{\begin{center}}
\newcommand{\EC}{\end{center}}
\newcommand{\BE}{\begin{equation}}
\newcommand{\EE}{\end{equation}}
\newcommand{\BEA}{\begin{eqnarray}}
\newcommand{\BEAnn}{\begin{eqnarray*}}
\newcommand{\EEA}{\end{eqnarray}}
\newcommand{\EEAnn}{\end{eqnarray*}}

\newcommand{\VL}{\left( \begin{array}{c}}
\newcommand{\VR}{\end{array} \right)}
\newcommand{\MLv}{\left( \begin{array}{cccc}}
\newcommand{\MR}{\end{array} \right)}

\newcommand{\Xt}{X_{t}}

\newcommand{\Xc}{X_{c}}

\newcommand{\At}{A_{t}}
\newcommand{\Aq}{A_{q}}
\def\tb{\tan\beta}

\begin{document}

\title{{\small{2005 International Linear Collider Workshop - Stanford,
U.S.A.}}\\ 
\vspace{12pt}
$\MW$, $\sweff$ and $\mh$ in the NMFV MSSM} 

\author{S.~Pe\~naranda, S.~Heinemeyer}
\affiliation{CERN, TH Division, Dept.\ of Physics, CH-1211 Geneva 23,
  Switzerland}
\author{W. Hollik}
\affiliation{Max-Planck-Institut f\"ur Physik,
F\"ohringer Ring 6, D--80805 Munich, Germany}
\begin{abstract}
The effects of loop contributions to the electroweak precision observables
and 
$\mh$ in the MSSM with non-minimal flavour violation (NMFV) are analyzed,
including the mixing 
between third and second generation squarks.
The mixing-induced shift 
in $\MW$ can amount to $140 \mev$ and 
to $70 \times 10^{-5}$ in $\sweff$ for extreme values of squarks
mixing, allowing to set limits on the NMFV parameters.
The corrections for $\mh$ are usually small and 
can amount up to \order{5 \gev} for large flavour violation.
\vspace*{0.6cm}\\
Pre-print numbers: CERN--PH--TH/2005--101, MPP--2005--58.
\end{abstract}

\maketitle

\thispagestyle{fancy}

\section{INTRODUCTION}
An alternative way, as compared to the direct search for Supersymmetry
(SUSY)~\cite{susy} or Higgs
particles, is to probe SUSY via virtual effects of the 
additional non-standard particles to precision observables. 
This requires very high precision 
of the experimental results as well as of the theoretical predictions,
such as the prediction for
$\De r$ in the $\MW$--$\MZ$~interdependence 
and the effective leptonic weak mixing angle, $\sweff$.

Radiative corrections to the electroweak precision observables
within the Minimal Supersymmetric Standard Model (MSSM) have been
discussed at the \onel\ and \twol\ level in several papers~\cite{summary}. 
Moreover, the Higgs sector of the MSSM is considerably
affected by loop contributions, thus making $m_h$ yet another sensitive 
observable. Here we present specifically the 
impact of non-minimal flavour violation (NMFV) in the MSSM on 
both the electroweak precision observables and the 
lightest Higgs-boson mass $m_h$. Thereby, mixing in the scalar
top and charm sectors as well as in the scalar bottom and strange 
sectors ($\Stop/\Scha$ and $\Sbot/\Sstr$) is considered. 
An exhaustive analysis is given in~\cite{nos}.

Mixing between the third and
second generation squarks can be numerically significant due to the
involved third-generation Yukawa couplings, shown in \cite{NMFVestimate}. 
On the other hand, there are strong experimental constraints on squark
mixing involving the first generation, coming from data on 
$K^0$--$\bar K^0$ and $D^0$--$\bar D^0$~mixing~\cite{FirstGenMix}.
Therefore, mixing effects from
first-generation squarks are not considered in our analysis.  
The general up-squark off-diagonal mass matrix in the basis of 
$(\SchaL, \StopL, \SchaR, \StopR)$ is given by  
\BEA
\label{eq:massup}
M_{\tiu}^2 &=& 
\MLv 
M_{\tilde L_c}^2 & \De_{LL}^t & \mc \Xc  & \De_{LR}^t \\[.3em]
\De_{LL}^t & M_{\tilde L_t}^2 & \De_{RL}^t &\mt \Xt \\[.5em]
\mc \Xc & \De_{RL}^t & M_{\tilde R_c}^2 & \De_{RR}^t \\[.3em]
\De_{LR}^t & \mt \Xt & \De_{RR}^t & M_{\tilde R_t}^2 
\MR 
\EEA
with
\BEA
\vspace*{-0.5cm}
\label{eq:defMXt}
M_{\tilde L_q}^2 &=& M_{\tilde Q_q}^2 + \mq^2 + 
                     \CZb \MZ^2 (T_3^q - Q_q \sw^2) \nn \\
M_{\tilde R_q}^2 &=& M_{\tilde U_q}^2 + \mq^2 + 
                     \CZb \MZ^2 Q_q \sw^2 ~(q = t,c) \nn \\
X_q &=& \Aq - \mu (\tb)^{-2 T_3^q}
\EEA
where $m_q$, $Q_q$ and $T_3^q$ are the mass, electric charge and
weak isospin of the quark~$q$;  
$M_{\tilde Q_q}$, $M_{\tilde U_q}$ are the soft
SUSY-breaking parameters; $A_q$ are the trilinear
Higgs couplings to $\Stop$, $\Scha\,$; $\mu$ is the Higgsino mass
                     parameter and  $\tb = v_2/v_1$.
Furthermore, $M_{Z,W}$ are the $Z$ and $W$ boson masses, and 
$s_W = \sin\theta_W$, $c_W = \cos\theta_W$ 
the electroweak mixing angle. Similar formula can be generated
for the $\Sbot/\Sstr$ sector by replacing $t\leftrightarrow b,
c\leftrightarrow s$. 

The numerical discussion of NMFV effects
and the illustration of the behaviour of $m_h$ and electroweak observables
are performed not for the most general case, but for the simpler and  
well motivated scenario where only mixing between the left-handed components
of $\Stop/\Scha$  and  $\Sbot/\Sstr$ is considered. 
The only flavour off-diagonal entries 
in the squark-mass matrices are normalized according to  
$\De_{LL}^{t,b} = \la^{t,b} M_{\tilde Q_3} M_{\tilde Q_2}$,
following~\cite{NMFVestimate,FirstGenMix} (the parameters 
$\la^t$, $\la^b$ are denoted by $(\delta_{{\scriptsize{LL}}}^u)_{23}$,
$(\delta_{{\scriptsize{LL}}}^d)_{23}$ in the above papers), 
where $M_{\tilde Q_3, \tilde Q_2}$ are the soft SUSY-breaking
masses for the $SU(2)$ squark doublet in the third and second
generation. 

In detail, we have
$\De_{LL}^t = \la^t M_{\tilde L_t} M_{\tilde L_c}\,,$
$\De_{LR}^t = \De_{RL}^t = \De_{RR}^t = 0 \,,$ 
for the entries in the matrix (\ref{eq:massup}) and, correspondingly,
$\De_{LL}^b = \la^b M_{\tilde L_b} M_{\tilde L_s} \,, 
\De_{LR}^b = \De_{RL}^b = \De_{RR}^b = 0$ for the down-squark mass
matrix. NMFV is thus parametrized in terms of the dimensionless
quantities $\la^t$ and $\la^b$. For the sake of simplicity, we have
assumed in the numerical analysis the same flavour mixing parameter in
the $\tilde t - \tilde c$ and $\tilde b - \tilde s$ sectors, 
$\la^t = \la^b = \la$. The case of $\la^t = \la^b = 0$ 
corresponds to the MSSM with minimal flavour violation (MFV). 

In order to diagonalize the two $4 \times 4$~squark mass matrices, two 
$4 \times 4$~rotation matrices, $R_{\tiu}$ and $R_{\tid}$, one for the
up-type squarks and the other one for the down-type squarks, are
needed. Once the squark mass matrices are diagonalized, one obtains the
squark mass eigenvalues and eigenstates that obviously depend on the flavour
mixing parameter $\la$. 
One important consequence of flavour mixing through the 
flavour non-diagonal entries in the squark mass 
matrices is generating large splittings 
between the squark-mass eigenvalues. For $\tilde t$ and $\tilde b$
squarks, one of the eigenvalues increases with $\la$ and the other one
decreases with $\la$ generating a large mass-splitting. For example, 
in the $\tilde b$-sector, the mass-splitting between $\tilde b_1$ and
$\tilde b_2$ is around $700 \gev$, with all the SUSY mass parameters fixed
to $1 \TeV$ and $\tb = 40$~\cite{HdecNMFV2}.
For more details and a list of Feynman rules for the restructured vertices,
i.e.\ the interaction of one
and two Higgs or gauge bosons with two squarks, we refer the reader
to~\cite{nos}. 

\section{$\De\rho$ AND ELECTROWEAK PRECISION OBSERVABLES}
\psfrag{lam}{{\scriptsize $\la$}}
\psfrag{lamu}{{\scriptsize $\la _{\tiu}$}}
\psfrag{lamd}{{\scriptsize $\la _{\tid}$}}
\psfrag{lamt}{{\scriptsize $\la_{t}$}}
\psfrag{lamb}{{\scriptsize $\la_{b}$}}
\psfrag{dr}{{\scriptsize $\De \rho $}}
\psfrag{MSusy}{{\scriptsize $M_{\rm {\tiny{SUSY}}} $ [GeV]}}
\psfrag{mhmax}{{\scriptsize $m_{h}^{\mbox{\tiny{max}}}$ scenario}}
\psfrag{nomix}{{\scriptsize no-mixing scenario }}
\psfrag{mhmax1000}{{\scriptsize $m_{h}^{\mbox{\tiny{max}}},
M_{\mbox{\fontsize{1}{1}\selectfont SUSY \normalfont}}=1 \tev $}}
\psfrag{nomix1000}{{\scriptsize no-mixing,
$M_{\mbox{\fontsize{1}{1}\selectfont SUSY \normalfont}}=1 \tev $}}
\psfrag{mhmax2000}{{\scriptsize $m_{h}^{\mbox{\tiny{max}}},
M_{\mbox{\fontsize{1}{1}\selectfont SUSY \normalfont}}=2 \tev $}}
\psfrag{nomix2000}{{\scriptsize no-mixing,
$M_{\mbox{\fontsize{1}{1}\selectfont SUSY \normalfont}}=2 \tev $}}
\psfrag{nomix}{{\scriptsize no-mixing scenario }}
\psfrag{dmw}{\rotatebox{90}{\footnotesize $\de \MW \, [GeV]$}}
\psfrag{dst}{\rotatebox{90}{\hspace*{0.5cm}\footnotesize $\de \sweff$  }}
The loop contribution to the electroweak $\rho$ parameter,
$\De\rho = \frac{\Si_Z(0)}{\MZ^2} - \frac{\Si_W(0)}{\MW^2}$, 
with the unrenormalized $Z$ and $W$ boson self-energies at zero momentum,
$\Si_{Z,W}(0)$,
represents the leading universal corrections to the electroweak
precision observables induced by
mass splitting between partners in isospin doublets~\cite{rho}
and is thus sensitive to the mass-splitting effects induced
by non-minimal flavour mixing. Precisely measured observables~\cite{ewdataw03}
like the $W$ boson mass, $\MW$, and the effective leptonic mixing 
angle, $\sweff$, are affected by shifts according to
\BE
\de\MW \approx \frac{\MW}{2}\frac{\cw^2}{\cw^2 - \sw^2} \De\rho, \quad
\de\sweff \approx - \frac{\cw^2 \sw^2}{\cw^2 - \sw^2} \De\rho .
\label{precobs}
\end{equation}

Beyond the $\De\rho$ approximation, the shifts in $\MW$ and $\sweff$
originate from the complete squark contributions to the quantity
$\De r$  and to other combinations of the various vector boson
self energies. However, we have numerically
verified that $\De\rho$ yields an excellent approximation for the full
calculation in the case of NMFV effects~\cite{nos}. 
The analytical \onel\ result for $\De \rho$, 
resulting from squark-loops based on the general $4 \times 4$~mass matrix 
for both the $\Stop/\Scha$ and the $\Sbot/\Sstr$ sector, 
has been implemented into the Fortran code \fehto~\cite{feynhiggs}.

For the numerical evaluation, the $\mhmax$ 
and the no-mixing scenario~\cite{LHbenchmark} have been selected, 
but with a free scale $\msusy$. 
In the $\mhmax$ benchmark scenario the trilinear coupling $\At$ 
is not a free parameter, obeying $\Xt= 2 \msusy$ $(\Xt=\At-\mu\cot\beta)$.
In the no-mixing scenario, $\At$ is defined by the requirement $\Xt=0$.
The results are independent of $\MA$.
The numerical values of the SUSY parameter are
$\msusy = 1\tev \; {\rm and}\; 2 \tev, \,\, \tb = 30, \,\, \mu=200 \gev\,$. 
However, the variation with $\mu$ and $\tb$ is very weak,
since they do not enter the squark couplings to the vector bosons. 

\begin{figure}[htb!]
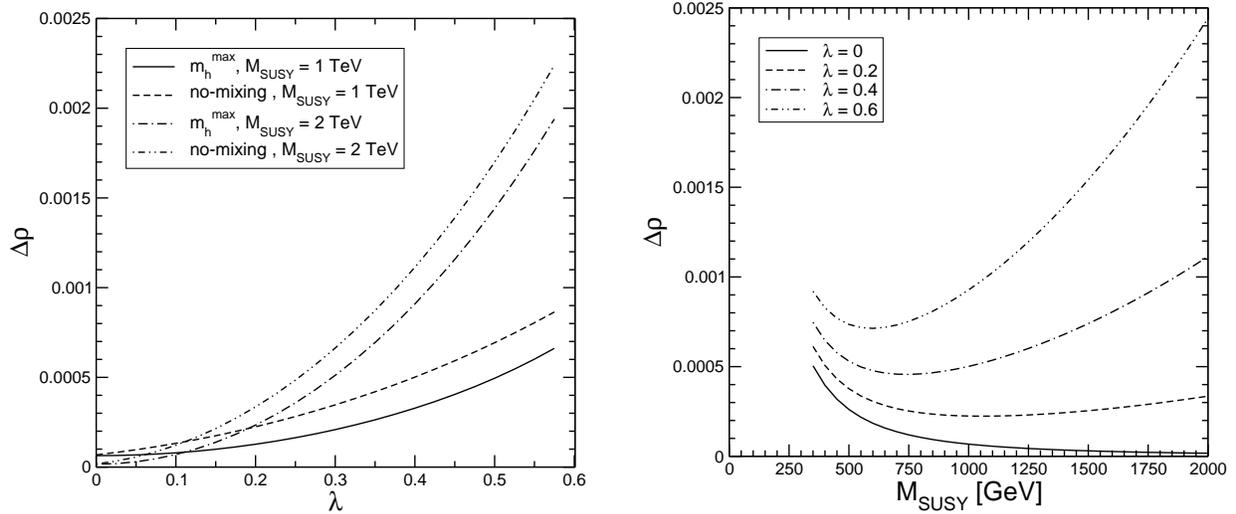

\begin{center}\hspace*{-0.5cm}
\epsfig{figure=deltarhol.eps,width=7.7cm}
\vspace*{0.3cm}\hspace*{0.5cm}
\epsfig{figure=drhoMsusy_nomix.eps,width=7.8cm}
\end{center}\vspace*{-0.8cm}
\caption[]{
The variation of $\De\rho$: with $\la = \la^t = \la^b$ in the
$\mhmax$~scenario and no-mixing~scenario, for 
$\msusy = 1 \tev$ and $2 \tev$ (left panel); and with $\msusy$ in the 
no-mixing~scenario for different values of $\la$ (right panel).}
\label{fig:rholam}
\end{figure}

In \reffi{fig:rholam} (left panel) we show the results for $\De\rho$ as
function of  $\la$ for both, the~$\mhmax$ and
no-mixing~scenario for two values of the SUSY mass scale. It is clear that
$\De\rho$ grows with the $\la$ parameter since the
splitting in the squark sector increases. One can also see that the
effects are close to zero for $\la = 0$ and $\msusy=2 \tev$, i.e we
recover the expected decoupling for $\la = 0$. 
For large values of $\msusy$ the contribution increases since the
splitting in the squark sector increases as well. 
The dependence on the SUSY mass scale is shown 
explicitely in \reffi{fig:rholam} (right panel) 
in the no-mixing~scenario, where the
effects on $\De\rho$ are largest.  
The region below $\msusy \lsim 400 \gev$ (depending on the
scenario) implies too low and hence forbidden values for the
squark masses. Correspondingly, for the largest values of $\la$ the
excluded region is larger.
For $\la = 0$ $\De\rho^{\tiq}$ decreases, being zero
for large $\msusy$ values as expected. Notice that the experimental
bound on $\De\rho$, $\De\rho\lsim 2\times 10^{-3}$, can be saturated. 

\begin{figure}[htb!]
\begin{center}
\vspace*{0.1cm}
\hspace*{-1cm}
\epsfig{figure=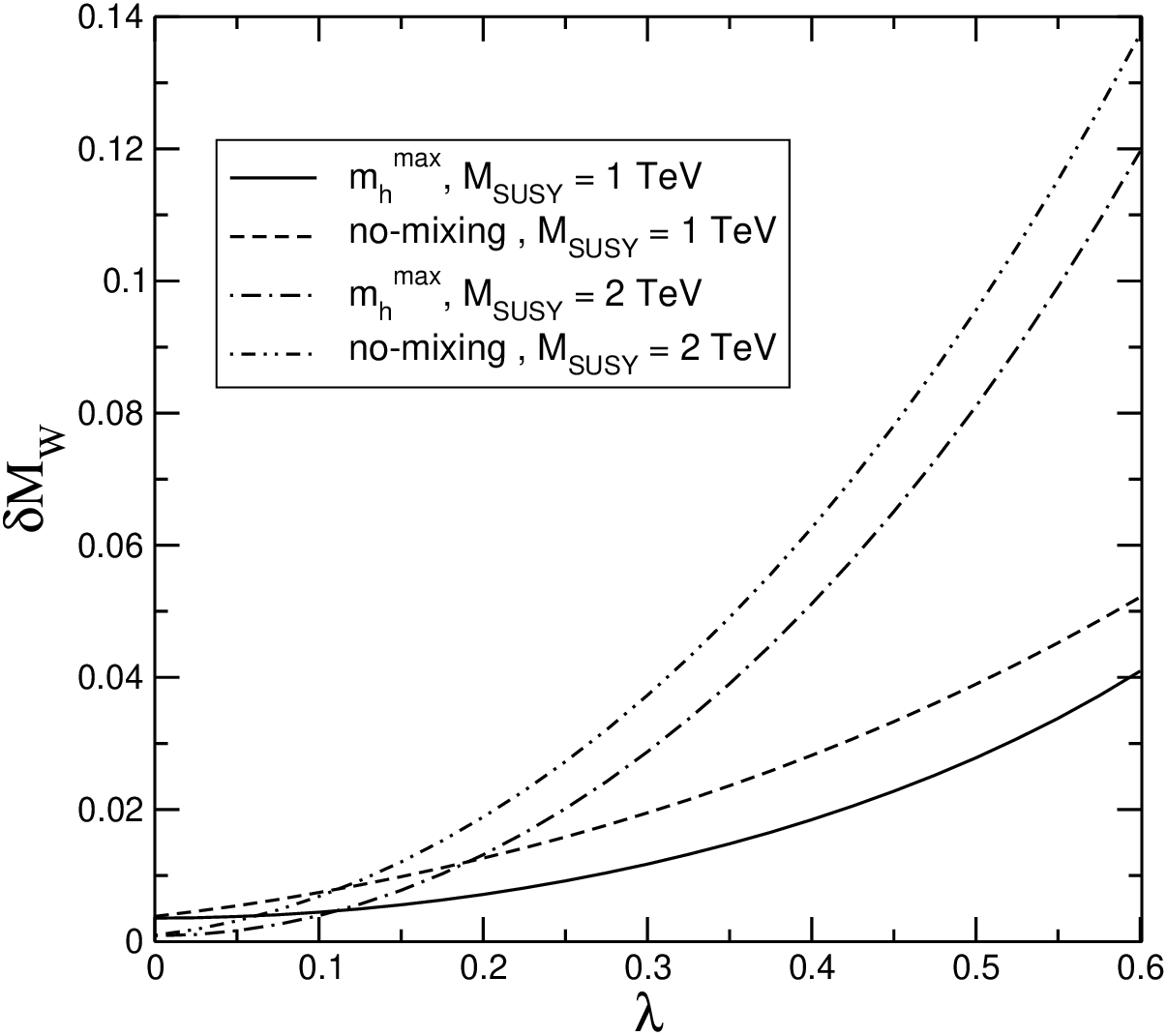,width=7.7cm}\hspace*{0.4cm}
\epsfig{figure=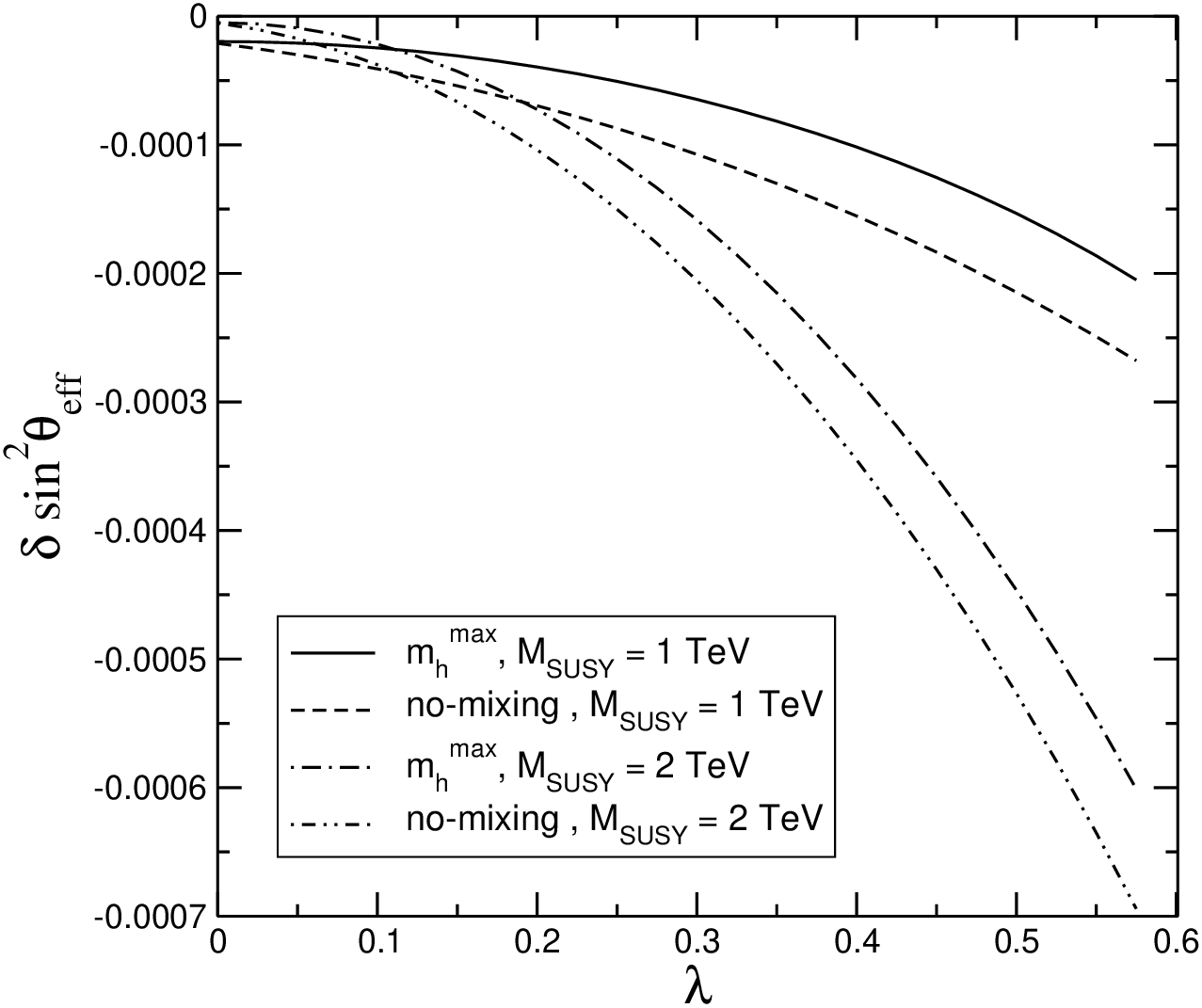,width=8cm}
\end{center}\vspace*{-0.6cm}
\caption[]{
The variation of $\de \MW$ and $\de \sweff$ as a function of 
$\la = \la^t = \la^b$ for the $\mhmax$ and
no-mixing~scenarios and different choices of $\msusy$. }
\label{fig:Ewprecision}
\end{figure}

The numerical effects of the NMFV MSSM contributions on the electroweak
precision observables, $\de \MW$ and $\de \sweff$ as a function of 
$\la$ are presented in \reffi{fig:Ewprecision}.
The $\mhmax$~scenario and no-mixing~scenario are included in both plots
and we choose two values of $\msusy$ as before. 
In the left plot the shifts $\de \MW$ are given, which can be as
large as $0.14 \gev$ for the no-mixing scenario  for 
$\msusy=2 \tev$, $\la=0.6$.
In the $\mhmax$~scenario $\de\MW$ remains smaller, $\de \MW \lsim 0.05
\gev$, but still sizeable. The correction to $\de \sweff$ is
shown in the right plot of \reffi{fig:Ewprecision}. One can see
that the shifts $\de \sweff$ can reach values up  
$7 \times 10^{-4}$ for $\msusy=2$ TeV and $\la=0.6$ in the
no-mixing~scenario, being smaller (but still sizeable)
for the other scenarios chosen here. 
These effects have to be compared with the current experimental
uncertainties, $\de\MW^{\rm exp,today} = 34 \mev, \,\,
 \de\sweff^{\rm exp,today} = 16 \times 10^{-5}$~\cite{ewdataw03}, 
the expected experimental precision for the LHC, 
$\Delta \MW = 15-20 \mev$~\cite{MWatLHC}, and at a future linear
collider running on the $Z$ peak and the $WW$ threshold (GigaZ),
$\Delta\MW^{\rm exp,future} = 7 \mev,  \,\,
 \Delta \sweff^{\rm exp,future} = 
1.3 \times 10^{-5} ~$~\cite{moenig}.
Extreme parts of the NMFV MSSM parameter space
can be excluded already with todays precision. 

\section{THE MASS OF THE LIGHTEST HIGGS BOSON}
The higher-order corrected masses $m_h,\, m_H$ of the $\cp$-even 
neutral Higgs bosons $h,H$ correspond to the poles of the $h,H$-propagator 
matrix. The status of the available results for the self-energy 
contributions to this matrix has been summarized in~\cite{mhiggsAEC}. 
Within the MSSM with MFV, the dominant \onel\ contributions to these 
self-energies result from the Yukawa part of the theory (i.e.\
neglecting the gauge couplings); 
they are described  by loop diagrams involving 
third-generation quarks and squarks.
Within the MSSM with NMFV, the squark loops have to be modified
by introducing the generation-mixed squarks.
Here we restrict ourselves to the dominant Yukawa
contributions resulting from the top and $t/\Stop$ (and
$c/\Scha$) sector.
Corrections from $b$ and $b/\Sbot$ (and
$s/\Sstr$) could only be important for very large values of
$\tb$, $\tb \gsim \mt/\mb$, which we do not consider here.
The analytical result of the renormalized Higgs boson self-energies,
based on the general $4 \times 4$~structure of the $\Stop/\Scha$~mass
matrix, has then been
implemented into the Fortran code \fehto~\cite{feynhiggs} that
includes all existing higher-order corrections (of the MFV MSSM). 

The numerical results for the lightest MSSM Higgs boson mass, $m_h$,
are presented for five benchmark scenarios
named ``$\mhmax$'' (to maximize the lightest Higgs boson mass), 
``constrained $\mhmax$'' (labeled as ``$\Xt/\msusy = -2$''),
``no-mixing'' (with no mixing in the MFV $\tilde t$ sector), 
``gluophobic Higgs'' (with reduced $ggh$ coupling), 
and ``small~$\aeff$'' scenario (with reduced $h b \bar b$ and $h \tau^+
\tau^-$ couplings)~\cite{LHbenchmark}.
For all these benchmark scenarios the soft SUSY-breaking parameters in
the three generations of scalar quarks are equal,
$\msusy = M_{\tilde Q_q} = M_{\tilde U_q} =M_{\tilde D_q}~$,
as well as  all the trilinear couplings, $A_s=A_b=A_c=A_t$. 

\begin{figure}[htb!]
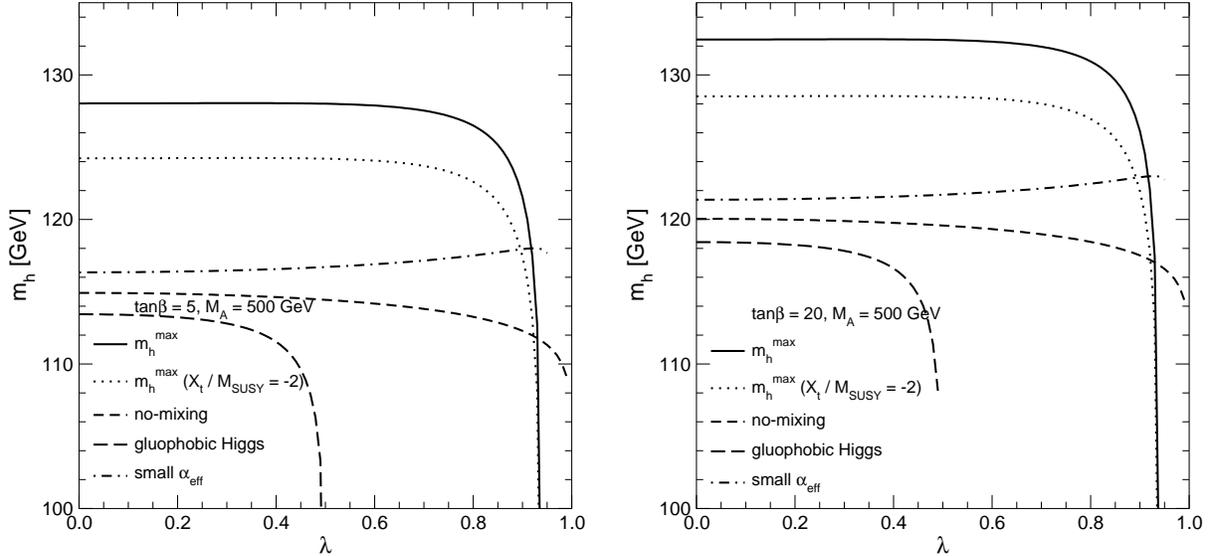

\begin{center}\vspace*{0.1cm}
\epsfig{figure=mhlam03.bw.eps,width=7.7cm}\hspace*{0.4cm}
\epsfig{figure=mhlam01.bw.eps,width=7.7cm}
\end{center}\vspace*{-0.4cm}
\caption[]{
The variation of $\mh$ with $\la = \la^t$ is shown in five
benchmark scenarios~\cite{LHbenchmark}.
$\MA$ has been fixed to $\MA = 500 \gev$, and $\tb$ is set to 
$\tb = 5$ (left panel) or $\tb = 20$ (right panel). }
\label{fig:mhmax}
\end{figure}

In \reffi{fig:mhmax} we illustrate the dependence of
$\mh$ on $\la( = \la^t)$ in all five benchmark scenarios.
$\MA = 500 \gev$ and $\tb$ is set to 
$\tb = 5$ (left) or $\tb = 20$ (right). 
All scenarios show a similar behavior. For small to moderate allowed
values of~$\la$ the variation of $\mh$ is small. Only for large values
(around 0.5 in the gluophobic Higgs scenario, and around 0.9 in
the other four scenarios) the variation of $\mh$ can be quite strong,
up to the \order{5 \gev}. In the gluophobic Higgs scenario unphysical
values for the scalar quark masses are reached already for smaller
values of~$\la$, since $\msusy$ is quite low in this scenario (see
\cite{LHbenchmark} for details). 
Values of $\lambda$ above $0.5$ imply forbidden values for the
squark masses in this scenario. In all cases except for the
small~$\aeff$ scenario the lightest Higgs boson mass turns out to be
reduced. In the small~$\aeff$ scenario it can be enhanced by up to $2 \gev$. 
Therefore, the impact of NMFV on $\mh$ is in general rather small, being at 
present lower than the theoretical uncertainty of $\mh$, 
$\de\mh^{\rm theo} \approx 3 \gev$~\cite{mhiggsAEC}.
Moreover, independent of low-energy FCNC data on flavour mixing, 
high values of $\la$ can be constrained by 
the experimental lower bound on $\mh$~\cite{LEPHiggs}.

Results presented in this paper have been later reproduced by using the
\fea, \foc \, packages~\cite{feynarts}, with a model file where the 
Feynman rules for the general NMFV MSSM (i.e.~including the generalized 
$6 \times 6$ squark mixing) have recently been implemented.

\section{CONCLUSIONS}
We have evaluated the scalar-quark contributions to
the lightest MSSM Higgs boson mass, to the $\rho$-parameter and to the 
electroweak precision observables $\MW$ and $\sweff$, arising from a
NMFV mixing between the third and second generation squarks.
The analytical results have been obtained for a general 
$4 \times 4$ mixing in the $\Stop/\Scha$ as well as in the
$\Sbot/\Sstr$ sector. They have been included in the Fortran code \fehto. 
The numerical analysis has been performed for a
simplified model in which only the left handed squarks receive an
additional mixing contribution, parametrized by
$\la$ (resp.\ $(\delta_{LL})_{23}$). 

Numerically we compared the effects of NMFV on the mass of
the lightest MSSM Higgs boson in five benchmark scenarios. For
small and moderate NMFV the effect is small. 
We have presented the numerical results for the squark contribution 
to the $\rho$-parameter. The additional contribution can be of
\order{10^{-3}}. Moreover, we have checked that even larger contributions can 
be obtained if the mixing in the
$\Stop/\Scha$ and $\Sbot/\Sstr$ sector is varied independently.
We have also analyzed the NMFV MSSM corrections to the electroweak
precision observables $\MW$ and $\sweff$, and we conclude that 
extreme parts of the NMFV MSSM parameter space 
can be excluded already with todays experimental precision of these
observables, and even more for the increasing precision at future
colliders. Therefore, scenarios analyzed in the context of  
B-physics can now be
  tested, whether they are compatible with electroweak precision
  observables (where the effects are large) and with Higgs physics
  (where the effects are small).

\begin{acknowledgments}
The work of S.P. has been supported by the European Community
under contract No. MEIF-CT-2003-500030. 
\end{acknowledgments}


\end{document}